# Stickiness in Chaos

G.Contopoulos and M. Harsoula
Research Center for Astronomy
Academy of Athens, Soranou Efesiou 4, 11527 Athens

**Abstract**

We distinguish two types of stickiness in systems of two degrees of freedom (a) stickiness around an island of stability and (b) stickiness in chaos, along the unstable asymptotic curves of unstable periodic orbits. In fact there are asymptotic curves of unstable orbits near the outer boundary of an island that remain close to the island for some time, and then extend to large distances into the surrounding chaotic sea. But later the asymptotic curves return close to the island and contribute to the overall stickiness that produces dark regions around the islands and dark lines extending far from the islands. We studied these effects in the standard map with a rather large nonlinearity K=5, and we emphasized the role of the asymptotic curves U, S from the central orbit O (x=0.5, y=0), that surround two large islands $O_1$ and $O'_1$, and the asymptotic curves $U_+U_-S_+S_-$ from the simplest unstable orbit around the island $O_1$. This is the orbit 4/9 that has 9 points around the island $O_1$ and 9 more points around the symmetric island $O'_1$. The asymptotic curves produce stickiness in the positive time direction (U,$U_+$,$U_-$) and in the negative time direction (S,$S_+$,$S_-$). The asymptotic curves $U_+$,$S_+$ are closer to the island $O_1$ and make many oscillations before reaching the chaotic sea. The curves $U_-$,$S_-$ are further away from the island $O_1$ and escape faster. Nevertheless all curves return many times close to $O_1$ and contribute to the stickiness near this island. The overall stickiness effects of $U_+$,$U_-$ are very similar and the stickiness effects along $S_+$,$S_-$ are also very similar. However the stickiness in the forward time direction, along $U_+$,$U_-$, is very different from the stickiness in the opposite time direction along $S_+$,$S_-$. We calculated the finite time LCN (Lyapunov characteristic number) $\chi(t)$, which is initially smaller for $U_+$,$S_+$ than for $U_-$,$S_-$. However after a long time all the values of $\chi(t)$ in the chaotic zone approach the same final value LCN=$\lim_{t \to \infty} \chi(t)$. The stretching number (LCN for one iteration only) varies along an asymptotic curve going through minima at the turning points of the asymptotic curve. We calculated the escape times (initial stickiness times) for many initial points outside but close to the island $O_1$. The lines that separate the regions of the fast from the slow escape time follow the shape of the asymptotic curves $S_+$,$S_-$. We explained this phenomenon by noting that lines close to $S_+$ on its inner side (closer to $O_1$) approach a point of the orbit 4/9, say $P_1$, and then follow the oscillations of the asymptotic curve $U_+$, and escape after a rather long time, while the curves outside $S_+$ after their approach to $P_1$ follow the shape of the asymptotic curves $U_-$ and escape fast into the chaotic sea. All these curves return near the original arcs of $U_+$,$U_-$ and contribute to the overall stickiness close to $U_+$,$U_-$. The isodensity curves follow the shape of the curves $U_+$,$U_-$ and the maxima of density are along $U_+$,$U_-$. For a rather long time the stickiness effects along $U_+$,$U_-$ are very pronounced. However after much longer times (about 1000 iterations) the overall stickiness effects are reduced and the distribution of points in the chaotic sea outside the islands tends to be uniform. The stickiness along the asymptotic curve U of the orbit O is very similar to the stickiness along the asymptotic curves $U_+$,$U_-$ of the orbit 4/9. This is related to the fact that the asymptotic curves of O and 4/9 are connected by heteroclinic orbits. However the main reason for this similarity is the fact that the asymptotic curves U, $U_+$, $U_-$ cannot intersect but follow each other.

**1.Introduction**

When we were trying to find the limits of some islands of stability in a dynamical system of two degrees of freedom we found some orbits that stayed for a long time around these islands, but then escaped to the large chaotic sea surrounding the islands (Contopoulos 1971). This was the first example of a stickiness phenomenon. Later similar results were found by many others (Shirts and Reinhardt 1982, Karney 1983, Meiss et al. 1983, Menjuk 1983,1985, etc). This phenomenon was called stickiness by Karney (1983).

It was realized that the stickiness phenomenon was due to the existence of cantori surrounding the sticky zone. Cantori are invariant sets consisting of infinite points, that do not form continuous lines (tori), but leave small gaps everywhere (Aubry 1978, Percival 1979). They are produced by the destruction of invariant tori, as the perturbation increases. When the holes are small they provide a partial barrier for the chaotic orbits inside them. Thus while the invariant curves do not allow any communication between an inner chaotic orbit with the outer chaos, when the last invariant curve (last KAM curve) is destroyed a communication of the inner and the outer chaotic regions is possible. Since that time much further work has been done on stickiness related to cantori (Chirikov and Shepelyansky 1984, Bensimon and Kadanoff 1984, MacKay et al. 1984, Harison et al. 1985, Meiss and Ott 1986, Meiss 1992, Mahon et al. 1995, Contopoulos et al. 1995, 1997, Kandrup et al. 1999, Contopoulos 2004 etc.). The phenomenon of the crossing of cantori was described in detail by Efthymiopoulos et al. (1997).

It is well known that chaos appears first near unstable periodic orbits. It is produced by the homoclinic and heteroclinic intersections of the asymptotic curves of unstable periodic orbits. However, up to now there has been no detailed study of the asymptotic curves that produce the effect of stickiness. It is the purpose of this paper to provide a detailed study of the asymptotic curves in a sticky region and their relation to the chaotic orbits in the large chaotic sea. The importance of such a study is due to the fact that the unstable periodic orbits and their asymptotic curves are different in different dynamical systems. Even completely chaotic systems (Anosov systems) have differences in the structure of their asymptotic curves. Thus the corresponding stickiness phenomena make such systems to look different for some time, even if later on all Anosov systems look similar.

In fact the stickiness effects sometimes last for very long times. E.g. in galactic dynamics they may last longer than the age of the Universe (Contopoulos 2004).

In the present paper we distinguish two types of stickiness:
(a) stickiness around an island of stability, and
(b) stickiness near unstable asymptotic curves in the chaotic sea. This is what we call "stickiness in chaos".

The stickiness around an island is produced by cantori surrounding this island, that do not allow a fast communication of the region inside a given cantorus with the region outside it. However stickiness appears also at relatively large distances from an island in the form of dark lines among a mostly uniform distribution of points in a chaotic sea (Fig.1). This stickiness is due to concentrations of points close to the unstable asymptotic curves of unstable periodic orbits. In fact the unstable asymptotic curves act as attractors of points outside them. Orbits close to a stable asymptotic curve, approach the unstable periodic orbit but later they deviate from the periodic orbit following closely unstable asymptotic curves. In particular if the eigenvalue λ of the periodic orbit is only slightly above 1 the nonperiodic orbits stay for a long time close to the periodic orbit (Contopoulos et al 1999). In general, the greater the eigenvalue λ of the periodic orbit, the faster the nonperiodic orbits will deviate from it.

The two types of stickiness are not clearly separated. In fact the sticky zone around an island follows the unstable asymptotic curves of the unstable periodic orbits in this region. These asymptotic curves pass later through the holes of the cantori and enter the large chaotic sea. Thus they produce stickiness of the second type, along particular dark lines that extend far from the islands in the chaotic sea. Such a type of stickiness appears even when the island disappears for a larger perturbation.

Similar stickiness phenomena appear close to the stable asymptotic curves if we change the direction of time. In the inverted time direction the roles of stable and unstable asymptotic curves are inverted. In the following we will use both directions of time in our calculations.

The concentration of points near unstable invariant curves is obvious only for some time. In fact after a long time the points deviate very much from each other and the darkness due to the distribution of points becomes practically everywhere equal to the darkness near the asymptotic curves, which are not conspicuous anymore (see section 6).

In the present paper we study orbits (sets of points) in the standard map

$$x_{i+1} = x_i + y_{i+1}$$
$$y_{i+1} = y_i + \frac{K}{2\pi}\sin(2\pi x_i)$$
(mod 1) (1)

with a time step $\Delta i=1$ (i is a kind of discrete time), for a value of K (K=5) where the stickiness phenomenon is evident. This value is relatively large in the sense that the phase space (x,y) contains very large chaotic domains, nevertheless there are still some important islands of stability. In the present case there are two main islands ($O_1$ and $O'_1$) symmetric with respect to the center (x=y=0.5), but most of the space around them is chaotic. The main (unstable) periodic orbits in the chaotic domain are the orbits (0,0) and O (0.5,0). The orbit (0,0) is unstable for all K>0. For small K (K>0) there is chaos practically only close to the asymptotic curves from (0,0). As K increases chaos increases and for K>0.97 the last invariant curve separating the chaotic domains close to y=1 from those close to y=0 is destroyed and chaotic orbits can go all the way from y=0 to y=1 (see, e.g. Contopoulos 2004).

The periodic orbit O (0.5,0) is stable for K<4 and unstable for K≥4. For K slightly larger than 4 a chaotic domain is generated near O but it does not communicate with the outer chaotic sea, until about K=4.35. However, for larger K the large chaotic sea reaches the region close to O and surrounds the islands around $O_1$ and $O'_1$. In all cases with K>4 the stickiness effects in the chaotic region are mainly regulated by the asymptotic curves of the simple periodic orbit O.

We study, first, a typical case of the asymptotic curves from this periodic orbit for a value of K=5 which is larger than K=4.35. In such a case the asymptotic curves from O surround the islands $O_1$, $O'_1$ and fill the whole chaotic sea outside these islands.

Around the main islands $O_1$ and $O'_1$ there are several unstable periodic orbits surrounded by cantori that allow communication with the outer chaotic sea. The simplest periodic orbit around $O_1$ for K=5 is the unstable orbit 4/9 that generates a large degree of stickiness, as we will see later. This orbit is generated from the orbit $O_1$ when K=4.8765 and its 9 points move outwards as K increases. For K slightly above K=4.8765 this orbit is inside a set of invariant curves around $O_1$, but for K=5 it is outside all invariant curves and its asymptotic curves reach the large chaotic sea. The last KAM curve that surrounds the orbit 4/9 for K<5 is the one with noble rotation number [2,3,1,1,…] (Efthymiopoulos et al, 1997), which has become a cantorus with large holes for K=5.

The periodic orbits, are found by giving an initial guess and then using a Newton-Raphson iterative method. Then we diagonalize the monodromy matrix

$$M = \begin{bmatrix} 1 + k\cos(2\pi x) & 1 \\ k\cos(2\pi x) & 1 \end{bmatrix} \qquad (2)$$

The eigenvalues satisfy the relation $\lambda_1 \cdot \lambda_2 = 1$ and the greater value corresponds to the unstable direction. In order to calculate the asymptotic curves we take a small initial segment of length $10^{-6}$ to $10^{-4}$ along the asymptotic vectors from the unstable periodic orbit and then we map a number of initial conditions on this segment forward in time along the unstable asymptotic curves and backward in time along the stable asymptotic curves. In order to produce dense asymptotic curves we vary the number of initial conditions empirically, taking a greater number for larger values of the eigenvalue of the unstable periodic orbit.

In the following sections we study the asymptotic curves of the orbit O (section 2), the asymptotic curves of the unstable orbit 4/9 (section 3), the Lyapunov characteristic numbers and the "stretching numbers" along the asymptotic curves (section 4), the escapes from the neighborhood of the island $O_1$ (section 5), the overall stickiness around the island $O_1$ (section 6) and the heteroclinic orbits (section 7). In section 8 we make some general remarks and draw our conclusions.

The asymptotic curves are described in some detail in order to emphasize the topological character of such curves. In other systems the details are different, but the topological features are common. Thus the present example should be considered as representative of much more general dynamical systems.

**2. Asymptotic curves of the unstable periodic orbit O**

For K=5 the orbit O (x=0.5, y=0) of period 1 is unstable, while the period-2 family $(O_1, O'_1)$, that bifurcated from O when K was equal to 4, is stable (Fig.2).

The asymptotic curves U (unstable) and S (stable) start at the point O upwards. (There are two more curves, symmetric to the above with respect to the center of the figure (x=y=0.5)).

The unstable asymptotic curve U after a maximum y, goes to a maximum deviation on the left from $O_1$ at the region U (Fig.2). Then it goes downwards to y=0 and beyond (continuing from y=1 because of the modulo 1) to the minimum A. Then it returns through y=1($\equiv$0) to a maximum y completing the thin lobe A. After that it forms a deaper lobe B and terminates at another maximum, larger than the previous one. Then it forms a long lobe C, that continues beyond a minimum C to the right and upwards. Further lobes are longer and approach closer and closer the original arc of the asymptotic curve U above O.

The stable asymptotic curve S is calculated backwards in time. Then we use the map:

$$x_{i-1} = x_i - y_i$$
$$y_{i-1} = y_i - \frac{K}{2\pi}\sin 2\pi(x_i - y_i) \qquad \text{(mod 1)} \qquad (3)$$

After a first maximum y the curve S goes through the region S on the right of $O_1$ (Fig.2) then downwards to y=0. The curve S continues from y=1 to a minimum of the lobe a, and then it goes to a maximum y, completing the thin lobe a. The next lobes are b, c, d,… each one longer than the previous one. These lobes pass closer and closer to the periodic orbit O, and to the original arc S. Successive lobes join at the maxima y: a-b, b-c, c-d (Fig.2). Every successive lobe is close to the previous one up to a certain distance, but then it deviates and extends to larger distances.

The asymptotic curves U,S from O (0.5,0) intersect the asymptotic curves of the unstable periodic orbit 4/9 at infinite heteroclinic points. Thus the curves U,S "guide" the asymptotic curves $S_\pm U_\pm$ from 4/9 and lead them into the chaotic sea (section 7).

### 3. Asymptotic curves of the unstable periodic orbit 4/9

This periodic orbit is represented by 9 points in Figs. 2 and 3a,b,c: $P_1, P_2, \ldots P_9$. There are 9 more points symmetric to the above with respect to the center (0.5,0.5) thus the orbit is in fact of multiplicity 18. The successive points in the mapping are $P_1, P_6, P_2, P_7, P_3, P_8, P_4, P_9, P_5$ i.e. every 4 points counterclockwise. Between the unstable points there are 9 points corresponding to a stable periodic orbit of period 18 (because there are 9 more stable points around $O'_1$). In Fig.3a we mark this stable periodic orbit by gray dots (that are surrounded by small islands of stability). We calculate the asymptotic curves: inner stable (slow) $S_+$ (Fig.3a), outer stable (fast) $S_-$ (Fig.3b) and inner unstable (slow) $U_+$ (Fig.3c), from every point $P_i$.

The calculations of Fig. 3a, b, c are made by plotting 8 successive iterates of $10^4$ evenly spread initial points along an eigenvector from every point $P_i$ within a distance $10^{-6}$ from $P_i$. We notice that Figs. 3a and 3b are similar, but Fig. 3b has many more points in the chaotic sea. This is due to the fact that particles along the fast stable asymptotic curves ($S_-$) escape faster in the chaotic sea and populate the chaotic sea for a longer time before returning close to the island $O_1$. In the same way the fast unstable asymptotic curve ($U_-$) is similar to $U_+$, but stays longer in the chaotic sea.

A comparison of the arcs $U_+$ and $S_+$ is shown in Fig.4. We see that the arcs $U_+$ and $S_+$ around the island 4/9 are quite different.

We describe now in some detail the various asymptotic curves. The arcs $S_+$, $U_+$ (slow) are inwards (towards the center of the island) while $S_-, U_-$ (fast) are outwards (Fig.5a). The point $P_1$ is at (x=0.642118, y=0.369344). This point is on the left of the islands 9/20, 14/31, 5/11,... that are described by Eftymiopoulos et al (1997). The resonance 4/9 is simpler than the other resonances (it is represented by fewer points). The islands 4/9, 9/20, 14/31 are in the chaotic zone outside the island $O_1$, while 5/11 is inside the last KAM curve surrounding the island $O_1$. The stable periodic orbit $O_1$ is further to the right, at (x=0.68002, y=0.36002).

The arcs $U_+, S_-$ (Fig.5a) intersect for the first time at a homoclinic point $h_1$, and define a resonance region $R_1$, while the arcs $S_+, U_-$ intersect at a homoclinic point $h'_1$ and define a resonance region $R_2$. The outer arcs $U_-$ and $S_-$ up to the homoclinic points are longer than the corresponding inner arcs $S_+$ and $U_+$.

### The asymptotic curve $S_+$

The inner asymptotic curve $S_+$ starts from $P_1$ upwards to the right (Fig.5a,b) towards the neighbourhood of $P_2$. It makes a number of oscillations close to $P_2$, going through the successive points $S_1, S_2, S_3, S_4, S_5, S_6, S_7$. One of the longest oscillations to the left reaches the minimum $S_8$ (Fig.5b), continues to the maximum $S_9$ then to the minimum $S_{10}$, maximum $S_{11}$ and minimum $S_{12}$. After that it returns along an almost identical parallel arc, to the maximum $S_{13}$, minimum $S_{14}$, maximum $S_{15}$, minimum $S_{16}$ and then outside the point $S_6$ upwards and to the right, to the neighborhood of $P_2$ and beyond it, up to the neighborhood of $P_3$ (point $S_{17}$ outside the figure). Then the curve $S_+$ returns to $S_{18}, S_{19}, S_{20}, S_{21}$ and proceeds downwards, all the way to the axis y=0 in the large chaotic sea.

After y=0 this arc continues, because of the modulo 1, from y=1 downwards (point $S_{22}$). After several further oscillations it comes again close to $S_{12}$, but to its right, then goes clearly above $S_{11}$ and $S_{10}$, through a maximum $S_{23}$ and a minimum $S_{24}$ and reaches $S_{25}$ (Fig. 5b). Then it returns, along a close by path, through $S_{26}$ and $S_{27}$ and reaches again the large chaotic sea going below the axis y=0 (point $S_{28}$). Thus a lobe is formed from the chaotic sea to the point $S_{25}$ and back to the chaotic sea. A similar lobe from the chaotic sea reaches the maximum $S_{29}$ (Fig. 5b) and returns to the chaotic sea (point $S_{30}$). Further lobes are formed above the point $S_{29}$ that fill a region between $P_9$ and $P_1$. Such are the lobes C and B of Fig.4. However a little above $S_{27}$ there is an island of stability around one point of the stable periodic orbit 4/9 that is avoided by the curve $S_+$. After a longer time the curve $S_+$ from $P_1$ comes close to the other periodic points $P_i$, but it cannot cross the corresponding curves $S_\pm$ from these points. In Fig. 3a we give the curves $S_+$ from all the points $P_i$.

The area A between and below the lines $S_8$-$S_9$ and $S_9$-$S_{10}$ is open and belongs to the large chaotic sea (Fig. 5b). This open area and the areas inside the lobes reaching the points $S_{25}$ and $S_{29}$ communicate, without any barrier from other arcs $S_+$, with the large chaotic sea. Therefore if an orbit starts inside the area A, or inside the lobes $S_{25}$, $S_{29}$ it escapes very fast.

The only uncertainty is what are the inner limits of the region A and of the lobes $S_{25}$, $S_{29}$ etc. If one calculates the asymptotic curve only up to the point $S_{12}$, one has the impression that the inner limit of the region A is provided by the arcs $S_8$-$S_9$-$S_{10}$-$S_{11}$-$S_{12}$. However after a longer calculation we find that the limit of A is marked by the arcs $S_{12}$-$S_{13}$-$S_{14}$-$S_{15}$-$S_{16}$ and after an even longer calculation we find that the limit of A is provided by the arcs $S_{18}$-$S_{19}$-$S_{20}$-$S_{21}$ going to $S_{22}$.

Now, are there further arcs of the asymptotic curve $S_+$ that restrict even further the area A? One expects the continuation of $S_+$ to fill the large chaotic sea itself, and this includes the area A and the lobes $S_{25}$, $S_{29}$. However this filling takes a very long time.

At any rate the asymptotic curve $S_+$ is probably ergodic outside the islands of stability, i.e. it probably approaches arbitrarily closely all points of the square (0,1)x(0,1) outside the islands.

A numerical argument in favor of this ergodic hypothesis is that the higher order iterations of the original points of the asymptotic curve $S_+$ are more or less smoothly distributed all over the large chaotic sea outside the islands $O_1$ and $O'_1$. This smooth distribution is more clearly seen in the iterates of the initial points along the curve $S_-$ (Fig.3b). Even inside the region A defined above there are scattered points (in Fig.3b) representing higher order images of points on the asymptotic curve $S_-$. Therefore it seems that there are no real empty regions outside the islands that are not reached by the curves $S_+$ and $S_-$.

**The asymptotic curve $S_-$**

While the asymptotic curve $S_+$ approaches $P_2$ and makes several oscillations around it, the curve $S_-$ (Fig.5a) goes directly into the large chaotic sea. The original arc $S_-$ from $P_1$ starts on the right and very close to the arc $S_{25}$-$S_{26}$ downwards, continuing above the point $S_{27}$ in Fig.5b, and then it goes fast to the axis y=0 between the lobes $S_{25}$ and $S_{29}$. Then from y=1 the curve $S_-$ goes to a minimum marked as "end $S_-$" inside the lobe b of S in Fig.2. Then it returns to y=1≡0 and from y=0 to a maximum y above y=0 and afterwards it goes again to the large chaotic sea. After going to large distances in the large chaotic sea, from time to time the curve $S_-$ comes back close to the periodic points $P_1$-$P_9$ (Fig.3b).

The overall distribution of the points of $S_-$ is similar to the distribution of the points of $S_+$ (Fig.3a). Although there are many differences in the details there are also many similarities. However we notice that the lines that are clearly seen in Fig.3a are approximated by fuzzy lines in Fig.3b because they correspond to higher order arcs where the distances between successive points are larger than during the first iterations. The points in the chaotic sea are more dense in Fig.3b than in Fig.3a. This means that S. spends more time in the chaotic sea than S$_+$.

The similarity of the overall forms of S. and S$_+$ is due to the fact that these curves cannot cross each other. Thus when the curve S. returns close to the island O$_1$ it has to avoid the various arcs of S$_+$ in this region, therefore it follows to some degree the arcs of S$_+$.

**The asymptotic curves U$_+$, U.**

The overall forms of the unstable asymptotic curves U$_+$ (Fig. 3c) and U. are similar to each other, but quite different from the stable asymptotic curves S$_+$, S. (Figs. 3a,b). However the first arcs of U. are rather different from those of U$_+$. The initial arc of U$_+$ from P$_1$ reaches the neighborhood of P$_9$ and makes oscillations near P$_9$ (Fig. 5a and c). After some oscillations of relatively short extent from maxima to minima y, e.g. 5, 6, 7 in Fig. 5c (the minima and maxima 1,2,3,4 are close to P$_9$ and are not marked in this figure) it reaches a minimum x (point 8 of Fig. 5c) on the left and after one more oscillation above the point 8 (points 9,10) it reaches a minimum y above P$_8$ (point 11). Then it goes above P$_1$P$_2$P$_3$, a little above the curve U., and to the right of P$_4$P$_5$P$_6$ downwards to the large chaotic sea close to the axis y=0 (Fig.5c). This happens at the 8$^{th}$ iteration of the original arc of length 10$^{-6}$ along U$_+$. Figure 5c giving the structure of the unstable asymptotic curve U$_+$, should be compared with Fig.5b which gives the structure of the stable asymptotic curve S$_+$. We notice that although U$_+$ makes several oscillations before going into the chaotic sea, its form is much simpler than S$_+$ which makes many more oscillations.

Similar lines extending to the large chaotic sea are generated from the other points P$_i$ of the periodic orbit 4/9.

After visiting many areas in the large chaotic sea the curve U$_+$ returns close to the points P$_i$ and forms new lobes, like the two lobes close and below P$_1$ in Fig.3c.

A comparison of the curves U$_+$ and S$_+$ in a region below P$_1$ is given in Fig.4. These curves surround an empty region occupied by an island around the stable orbit 4/9. The curves U$_+$ and S$_+$ have quite different forms and they intersect at an infinity of homoclinic points.

The asymptotic curve U. goes to large distances (Fig.5a) and reaches the large chaotic sea quite fast. In fact the 5$^{th}$ iteration of the original length 10$^{-6}$ along U. from P$_1$ not only reaches the axis y=0 but forms also several lines in the chaotic sea, surrounding the whole island O$_1$ (Fig.5c) and then enters again in the region close to P$_1$.

The overall form of U. is similar to U$_+$ (Fig.3c), but its lines are more fuzzy, because U. stays for a longer time in the large chaotic sea and it returns close to the points P$_1$,P$_2$…P$_9$ later. The similarity of U. and U$_+$ is due to the fact that these lines cannot cross each other, therefore when they return close to the island O$_1$ they form similar arcs around this island.

**4. Lyapunov characteristic numbers and stretching numbers**

The degree of stickiness is related to the Lyapunov characteristic number (LCN) and in particular with the "local LCN" (or "stretching number" Voglis and Contopoulos 1994) of the orbits

$$\alpha_t = \ln |\xi_{t+1}/\xi_t| \tag{3}$$

where $\xi_i$ is the deviation at time $i$. In general small stretching numbers generate a larger stickiness.

Very near an unstable periodic orbit the stretching number is equal to the logarithm of the larger eigenvalue of the orbit. The eigenvalue of the orbit $P_1$ is $\lambda_0=1.1546$ for one iteration, i.e. $\lambda=(1.1546)^{18} = 13.298$ after 18 iterations, i.e. one return close to $P_1$. The Lyapunov characteristic number (LCN) of the periodic orbit $P_1$ is equal to $LCN(P_1)=\ln\lambda=2.5876$. We take $10^4$ points along $S_+$, $U_+$, $S_-$, $U_-$ with a step $10^{-8}$ (total length along each arc $10^{-4}$). After n 18-ple iterations the length of an initial arc $L_0$ along $U_+$, $U_-$ is

$$L_n=L_0\lambda^n \qquad (4)$$

After 4 iterations the arcs along $S_+$, $U_+$, just reach the neighborhoods of $P_2$ and $P_9$ respectively after several oscillations (Fig.5a) while the arcs $S_-$, $U_-$ extend to longer distances, and at their 3$^{rd}$ iteration they reach the chaotic region near the axis $y=0$. After 5 iterations the initial length $10^{-4}$ along $S_+$ reaches approximately the point $S_6$ of Fig.5b and after 6 iterations it describes all the curves of Figs.5b (including the lobe $S_{29}$).

The average value of the stretching number is the "finite time LCN", defined by the relation

$$\chi(t) = \ln(\xi/\xi_0)/t = (1/t)\sum_{t=0}^{t_i} a_i \qquad (5)$$

where $\xi$ and $\xi_0$ are the deviations of two infinitesimally close orbits at times t and 0.
The function $\chi(t)$ for initial conditions close to $P_1$ along the manifolds $U_+$, $U_-$, $S_+$, $S_-$ is given in Fig. 6a for a relatively short time. After a much longer time the value of $\chi(t)$ tends to the usual "Lyapunov characteristic number" $LCN = \lim_{t\to\infty} \chi(t)$.

We see that $\chi(t)$ is almost constant up to $t\approx 6$ (Fig.6a), approximately equal to $\alpha=\ln(13.298)=2.588$, i.e. equal to the value of $a$ at $P_1$. The constancy of $a$ means simply that up to 6 iterations the points remain so close to $P_1$ that the linear theory is applicable there. After t=6 the values of $\chi(t)$ for the various asymptotic curves start to be differentiated. In fact if the initial length is $L_0=10^{-9}$ its 6$^{th}$ iteration is $L_6=0.0055$, i.e. a rather small length. The values of $\chi(t)$ are smaller for $S_+$, $U_+$ than for $S_-$, $U_-$, for some time, but after about $t\approx 10-15$ the values of $\chi(t)$ for $S_+$, $U_+$ grow rather abruptly. After about $t\approx 300$ the curves $\chi(t)$ tend to a fixed final value which is the Lyapunov characteristic number $LCN \approx 17.0$ of the large chaotic sea. In fact if we take orbits starting in the chaotic sea and arbitrary initial deviation $\xi_0 = (\xi_{xo}^2 + \xi_{yo}^2)^{1/2}$ we find the same limit $LCN \approx 17.0$ as above. An example is shown as a thick line C in Fig. 6a. The orbit C starts at (x=0.3, y=0.5) with infinitesimal deviations $\xi_{xo} = \xi_{yo}$, and the iterates $\xi_{xi}, \xi_{yi}$ are found from the variational equations corresponding to Eq.(1), but after 18 iterations in order to compare the deviations from this orbit with the deviations along $U_\pm, S_\pm$.

We notice that the LCN of the chaotic orbits is LCN~17.0 is much larger than the LCN of the periodic orbit $P_1$ (LCN=13.298 after 18 iterations).

In Fig. 6b we see that the value of the stretching number $\alpha$ along the asymptotic curve $U_+$ has a characteristic form. It starts at $\alpha =\ln(13.298)$ which is the LCN of the periodic orbit $P_1$ and decreases to a minimum. Then it goes from this minimum to a maximum and back to another minimum. This behavior is repeated indefinitely. The maxima increase in general (but not always), reaching values close to $α =17.0$ after many iterations. Thus we explain why LCN along an asymptotic curve is equal to the value of LCN of a chaotic orbit.

In order to explain the variations of $α$ along the asymptotic curve $U_+$ we should compare Fig.6b with the form of the curve $U_+$ (Fig 5c) where we mark its turning points. We find that the minima $α$ occur at these turning points, which are rather abrupt.

Another quantitative estimate of the effect of stickiness was provided recently by Skokos et al. (2007).

As regards the accuracy of our calculations we should note the following. As the eigenvalue of the orbit (after 18 iterations) is $λ=13.3$ the error becomes larger than 1 for $L_0=10^{-9}$ after $n=8$ iterations. Therefore after 8 iterations the details of an orbit cannot be trusted; only its general form may be considered.

On the other hand the accuracy of the lobes described above after 6 iterations are quite accurate. E.g. we found that the detailed descriptions of the asymptotic curves S, U, $S_±$, $U_±$, are the same even if we use single precision calculations. Only very close to the points $P_i$ we need a greater accuracy in our calculations.

## 5. Escapes

We define the "escape time" along any orbit starting around the island $O_1$ of stability as the number of iterations required for this orbit to reach the large chaotic sea. In our previous paper (Efthymiopoulos et al 1997) we considered an orbit as escaping if it went outside a given ellipse surrounding the island. In the present paper we consider orbits escaping when they cross the axis $y=0$. In view of the fact that the orbits escape downwards from the neighborhood of the island $O_1$ the two definitions do not give appreciably different results.

In that paper the escape time was called also "stickiness time". A more appropriate name is "initial stickiness time" because it refers only to the stickiness until the image of each point escapes to the chaotic sea. However later on images of every point return to the same region and contribute to the overall stickiness there (see section 6).

The escape times of orbits starting in a region on the left of the island $O_1$ are given in Fig. 7. We mark with different colors orbits escaping after 1-5 iterations (red), 5-10 iterations (yellow), 10-100 iterations (green), 100-1000 iterations (blue) and over 1000 iterations, or not escaping (gray on the right and close to 4/9). The main remark is that the regions of fast escapes are delineated by the stable asymptotic curves. In Fig. 7 we mark the elongated regions A, B, C of Fig. 4. We also mark the stable asymptotic curve $S_+$ from $P_1$ with black lines. We observe that the limits of the regions of fast escapes are separated from regions of slower escapes by arcs along the asymptotic curve $S_+$. The upper left part of Fig. 7 belongs to the large chaotic sea. The right part of the figure (gray) contains regions of very slow escapes, or regions belonging to the island $O_1$ with no escape at all. Close to the right boundary of the island $O_1$ we have stickiness around secondary islands (gray). In the lower part of the figure we see also a sticky region of the island 4/9 (gray) surrounding the stable periodic orbit 4/9.

An explanation of the role of the stable manifold $S_+$ in distinguishing between fast and slow escapes can be provided as follows. In Fig.8 we give the successive images of two lines, $K^+$ and $K^-$ very close to $S_+$ above the point $P_1$, at distances $Δx=±10^{-4}$ to the right and to the left of $S_+$. The initial lengths of $K^+$, $K^-$ extend approximately from $P_2$ to $P_1$. In the scale of this figure $K^+$ and $K^-$ practically coincide with $S_+$ above $P_1$.

The orbits starting close to the stable asymptotic curve $S_+$ have their images close to the same asymptotic curve until they reach the neighborhood of the periodic orbit $P_1$ and then they deviate close and along $U_+$ or $U_-$. Namely orbits starting along $K^+$, inside the curve $S_+$ (i.e. closer to $O_1$) deviate along a line close to $U_+$ (black line, Fig. 8) and points along $K^-$ outside $S_+$ deviate close to $U_-$ (red line, Fig. 8). Similarly, orbits starting inside $S_-$ deviate close to $U_+$ and orbits starting outside $S_-$ deviate close to $U_-$ (Fig.5a).

The curve $K^+$ approaches for some time the curve $U_+(P_1)$ from $P_1$ below $P_1$. In fact the successive points close and below $P_1$ approach $U_+$ along lines almost parallel to the stable asymptotic curve $S_+$ at a rate $1/\lambda$, where $\lambda > 1$ is the larger eigenvalue of $P_1$. However, further away from $P_1$ the successive points deviate from $U_+$. E.g. the curve $K^+$ makes a few oscillations similar to the oscillations of $U_+$ $(P_1)$ close to $P_9$ but these oscillations are with smaller amplitude than in $U_+$. Below $P_9$ the curve $K^+$ makes some oscillations close to $P_8$ and then close to $P_7$. After $K^+$ reaches the neighborhood of $P_8$ it deviates completely from the asymptotic curve $U_+(P_1)$, which does not come very close to $P_8$. The lowest point of $U_+(P_1)$ above $P_8$ is marked in Fig.8.

After reaching the neighborhood of $P_7$ the curve $K^+$ returns to the left and upwards from $P_8$, and makes some oscillations on the left of $P_9$. In Fig. 8 we mark the end of the $6^{th}$, $7^{th}$ and $8^{th}$ iterations. After some further oscillations beyond 8, the $9^{th}$ iteration of $K^+$ goes upwards, above the line 12-13 of Fig.5c, and then downwards close to the line 13 of Fig.5c reaching the large chaotic sea at $y=0$.

Thus we see that the curve $K^+$ reaches the chaotic sea, but after a large number of oscillations and a relatively long time.

On the other hand if we follow the iterations of a curve $K^-$ above and on the left of $S_+$ from $P_2$ to $P_1$ starting at initial distances $\Delta x=-10^4$ from $S_+$, this curve approaches the curve $U_-$ near and above $P_1$ and reaches the chaotic sea very soon (Fig.8). In fact the curve $K^-$ makes some oscillations near $P_2$, following the oscillations of $U_-$, but then it deviates further from $U_-$ and goes to the large chaotic sea near $y=0$ after only 3 iterations.

## 6. Overall Stickiness

The "escape time" is not sufficient to characterize the overall stickiness of an orbit. The reason is that even if an orbit escapes from the neighbourhood of the island $O_1$ it returns many times close to $O_1$ and remains for certain intervals of time close to the unstable manifolds of the orbit 4/9. In particular the unstable asymptotic curves, after exploring the large chaotic sea, return close to the original unstable asymptotic curves in the neighbourhood of the boundary of the island $O_1$, thus contributing to the overall stickiness around $O_1$. Although the escapes from the region around the island $O_1$ are governed by the stable manifolds of the orbit 4/9, the overall stickiness is governed by the unstable manifolds of the orbit 4/9 and also by the unstable manifold of the orbit O that surrounds the island $O_1$.

The stickiness close to the island $O_1$ can be seen in Fig.3c, where we have calculated the asymptotic curves $U_+(P_i)$ from the points $P_i$ (i=1,2,…9) starting with initial lengths $10^{-6}$ for 8 iterations. A very similar figure is provided by the asymptotic curves $U_-(P_i)$. Although the asymptotic curves $U_-(P_i)$ go to the chaotic sea faster than the asymptotic curves $U_+(P_i)$, they return many times close to their original lines and to the lines formed by the asymptotic curves $U_+$. The only difference between $U_+$ and $U_-$ as we have seen in section 3, is that the sticky regions are more fuzzy for $U_-$ and the chaotic domain is more dense than in Fig.3c, because the asymptotic curves $U_-(P_i)$ stay for a longer time in the chaotic sea.

The overall stickiness due to the asymptotic curve $U_+$ from $P_1$ is seen in Fig.9a, which gives 15 iterations of 20000 initial points in an interval $10^{-6}$ along the asymptotic curve $U_+(P_1)$. In this figure we see not only a very dark region around the island $O_1$, that represents stickiness of the first type, but also dark lines extending to relatively large distances, mainly downwards, to $y=0$, and continuing from $y=1$ downwards, surrounding also the island $O'_1$. Very similar figures are provided if we calculate the asymptotic curve $U_-(P_1)$ for an equal number of iterations, and also if we calculate all the asymptotic curves $U_+(P_i)$, or $U_-(P_i)$ with i=1,2,...9 for a long time.

If we calculate an equal number of iterates (15) from an equal number of initial points (20000) along an interval $10^{-6}$ from the periodic orbit O(0.5,0) (Fig. 9b), we find that the iterates do not spread all over the phase space. This is due to the fact that the eigenvalue of the orbit O ($|\lambda|=2.6$) is much smaller absolutely than the eigenvalue of the 4/9 type orbit $P_1$ ($|\lambda|=13.6$). Because of this difference of the eigenvalues the effect of the orbit $P_1$ on the dynamics of the system is much more pronounced than the effect of the orbit O.

In all three cases the higher order iterates along the asymptotic curves $U_+(P_1)$ (Fig. 9a) and U(O) (Fig. 9c) are separated according to the law $ds_n = |\lambda|^n ds_0$, therefore they do not form continuous lines. However it is clear that they fill the same space outside the islands of stability, and they are concentrated along very similar lines close and outside these islands. This is due to the fact that the asymptotic curves from the orbits O and $P_1$ are united into a common complex set because of their heteroclinic intersections (section 7).

Furthermore we note that Figs.9a,c are very similar to Fig.1, which represents 50 images of $10^4$ initial conditions along a line of constant y, although this figure does not represent orbits starting on any asymptotic curve. This similarity shows that the overall stickiness affects the whole map in the sense that orbits starting at various distances from the island $O_1$ have many images that surround closely this island (and to the island $O'_1$) and also close to the sticky lines produced by the asymptotic curves of the unstable orbits O and 4/9. Of course orbits starting very close to the island $O_1$ stay around this island for a longer time before filling more or less smoothly the large chaotic sea.

In order to estimate better quantitatively the overall stickiness around the island $O_1$ we have calculated, in Fig.10a, the distribution of 100 iterates of 500000 initial points in the same area of Fig.7 (which gives the escape times).

If we compare Fig.10a with Fig.4 we can see that the maximum density is close to the asymptotic curves $U_+$, while the density decreases away from these lines. The contrast between Fig.10a and Fig.7 is striking. It is clear that the maximum overall stickiness (which is along the lines $U_+$) is not along lines of large escape time (large "initial stickiness times", which are defined by the curves $S_+$).

This difference is striking, because one may think that when the orbits that have escaped into the large chaotic sea come back in the same area (the area of Figs 7 and 10a) outside the island $O_1$, they should follow the same pattern as that provided by the "initial stickiness times". In fact if an orbit returns to a particular point of this region its subsequent escape time is defined unambiguously by the color of this point in Fig.7. However, Fig. 7 implies that we have a homogeneous initial distribution of particles and the density of the particles decreases fast in time in the red regions on the left and less and less fast in the yellow, green, blue and gray regions. This figure does not give the density of particles that return to this area after their escape to the chaotic sea. On the other hand the distribution of the returning particles at any given time is far from homogeneous. The particles return after very different times, and the returning points are concentrated preferably close to the unstable asymptotic curves in that region. Thus, although the new escape times from any point are equal to the original escape times, the density of particles at any given time is maximum near the unstable asymptotic curves U and $U_+$, $U_-$ in this region, as in Fig. 10a, i.e. very different from the distribution in Fig.7.

There is one more point to be discussed in this section, namely how long does the overall stickiness last. If we calculate orbits for longer and longer times we see that they tend to populate more and more evenly the whole space of the map (0,1)x(0,1) outside the islands.

This is true even for orbits starting along the asymptotic curves of the periodic orbits O and 4/9.

The reduction of the stickiness after long times can be seen if we compare Fig.10a, that gives the distribution of the first 100 iterates of the 500000 initial points (in the area of the figure), with Figs 10b and 10c that give the distribution of 100 iterates between t=400-500 (Fig.10b) and between t=1000-1100 (Fig.10c). We see that the iterates are spread more evenly in Fig.10b and even more evenly in Fig.10c. In Fig.10b we still see stickiness close to the unstable manifold $U_+$ of the unstable orbit 4/9 but in Fig.10c the stickiness is restricted only close to the boundaries of the island $O_1$ (right part of the figure) and close to the stable island 4/9 in the lower boundary of the figure.

The overall tendency to smoothness is seen in Fig.11, that contains 20 iterates of the same 2500 initial points of Fig. 9a , but after a time T=1000, and distributed in the whole space (0,1)x(0,1). We see that the density of the points outside the islands is practically constant and no stickiness effect is apparent. This is seen very clearly if we compare Fig.11 with Fig.9a. However very close to the islands there are still some very small stickiness regions due to higher order unstable periodic orbits around the islands.

As a conclusion we find that the stickiness effects around a relatively large neighbourhood of the island $O_1$ (as seen in Figs.1, 9a,b and 10a,b,c) are important only for some hundreds of iterations. After 1000 iterations only the close neighbourhood of the various islands still shows some stickiness effects. At that time the dark lines that extend far into the chaotic sea (Figs. 1, 9a,b) fade and most of the total area of the map (i.e. (0,1)x(0,1) outside the islands) tends to be evenly populated by the iterates of the various orbits.

The total number of points after 1000 iterations is 500000x1000=$5 \times 10^8$. This is consistent with the results of a previous calculation (Contopoulos et al. 1995) where we have found that the distribution of $10^8$ iterates of one point in the chaotic sea is almost uniform outside the islands. Thus $10^8$ iterates is a rough estimate of the stickiness time.

Outside the last KAM curve around the island $O_1$ there are many unstable orbits that communicate with the large chaotic sea. The most important among them are the orbits 9/20 and 14/31. These orbits are at the centers of the largest islands close to the outer boundary of the orbit $O_1$. Their unstable asymptotic curves intersect the stable asymptotic curves of the orbits 4/9 and O(0.5,0) thus they belong to the same complex of orbits. Their stickiness is similar to that of the orbit 4/9 although there are some small differences in the inner parts.

**7. Heteroclinic intersections**

The stickiness due to the asymptotic curves of various unstable orbits around the island $O_1$ is similar because these asymptotic curves intersect each other. E.g. the unstable asymptotic curve U of the orbit O (0.5,0) intersects the asymptotic curves $S_\mp$ of the orbit 4/9. Similarly the stable asymptotic curve S of O intersects the unstable asymptotic curves $U_\mp$ of the orbit 4/9.

The intersections are heteroclinic points forming heteroclinic orbits that approach the orbits O and 4/9 asymptotically as the number of iterations in the forward and backward time direction tends to infinity.

An example is shown in Fig.12, where we give a few intersections of the curves U and $S_-$. The initial arcs of the curve U of Fig.2 are shown as thick dark lines in Fig.12. This curve starts at O and surrounds the island $O_1$ (Fig.2). In Fig.12 the curve U comes from above, makes two oscillations on the left, and then it intersects the curve $S_-$, which starts at the point $P_1$, at two heteroclinic points $H_1$ and $H_2$. The segment $H_1H_2$ is almost a straight line along $S_-$, but it forms a lobe to the right of this straight line along U.

The first image of the segment $H_1H_2$ along $S_-$ (forward in time) is a very small segment $H'_1,H'_2$ close to $P_1$. However the image of the lobe $H_1UH_2S_-H_1$ is an extremely long lobe that goes all over the whole square (0,1)x(0,1). The reason is that one iteration along $S_-$ corresponds to 18 iterations of the map (1), therefore the arc $H_1UH_2$ is mapped into a very long arc $H'_1UH'_2$. This arc starts at $H'_1$ downwards and then to the left, and after a very large number of oscillations it reaches a remote point (outside Fig.12) and then returns to the point $H'_2$ a little below $H'_1$. The returning arc is usually very close to the original arc but close to angular points we see double lines that differentiate the original arc from the returning arc. The area of the thin lobe $H'_1UH'_2S_-H'_1$ is equal to the area of the original lobe $H_1UH_2S_-H_1$.

If we continue the mapping, the successive points $H_1H'_1H''_1\ldots$, and $H_2H'_2H''_2\ldots$, come very close to $P_1$ approaching it asymptotically along $S_-$. At the same time the images of the arc $H_1UH_2$ are much longer covering coarsely the whole area of the square $(0,1)\times(0,1)$ outside the islands.

The heteroclinic points connecting U with $S_-$ and $S_+$, and also S with $U_-$ and $U_+$, anchor the manifolds of the orbits O and 4/9 to each other and form a joint manifold. There are also heteroclinic intersections between the orbits O, 4/9 and the orbits 9/20 and 14/31. On the other hand the orbit 5/11 is inside the last KAM curve around the island $O_1$ and has no heteroclinic intersections with the orbits O, 4/9 etc. In fact the non existence of heteroclinic points is a proof that the orbit 5/11 is separated from the orbits O, 4/9 by a KAM curve. Thus the chaotic regions around the orbit 5/11 are very different from the chaotic regions around the orbits O and 4/9.

The form of the joint manifold of the asymptotic curves O and 4/9 is governed mainly by the lack of intersections between U and $U_\mp$, and between S and $S_\mp$. This lack of intersections forces the unstable manifolds U, $U_+$ and $U_-$ to follow each other and be very similar in general, despite their differences in the details. Similarly the stable manifolds S, $S_+$ and $S_-$ are similar to each other. This fact explains the similarity of the stickiness due to the orbits O and 4/9.

## 8. General remarks and conclusions

Stickiness is an important characteristic of dynamical systems that have both order and chaos. Sticky orbits stay for a long time in certain regions of phase space producing an uneven distribution of points in the chaotic zone. The stickiness effects may last for very long times, longer than the age of the Universe in certain cases.

Up to now most of the work on stickiness has been restricted to stickiness around islands of stability. Such stickiness is due to the existence of cantori, i.e. destroyed KAM curves surrounding an island of stability. If the holes of such a cantorus are small, chaotic orbits starting inside this cantorus take a relatively long time to cross the cantorus and reach the large chaotic sea outside the cantorus. However there is also a second type of stickiness, namely stickiness along the unstable asymptotic curves, which extend to large distances in the chaotic sea. Thus we see in the chaotic domain not only concentrations of points around the islands of stability, but also dark lines that extend far into the chaotic sea.

The two types of stickiness are connected. In fact the asymptotic curves of the unstable orbits in the sticky zone around an island continue beyond the cantori surrounding these unstable orbits, and form dark lines extending quite far from the island.

In the present paper we studied stickiness in the standard map for a particular value of the perturbation parameter (K=5) when we have two main islands of stability, around two stable orbits $O_1$ and $O'_1$, that have bifurcated from a central periodic orbit O (x=0.5, y=0) when K increased beyond K=4. The islands of stability are surrounded by several periodic orbits that have bifurcated from $O_1$ and $O'_1$.

We considered mainly the unstable periodic orbit 4/9 that produces an important stickiness domain around the islands $O_1$, $O'_1$. However this domain communicates also with the large chaotic sea that covers most of the space $(0,1)\times(0,1)$ outside the islands.

The main conclusions of our study are the following:

(1) There are two types of stickiness. The first is due to cantori, surrounding islands of stability. The second is along the unstable asymptotic curves of the unstable periodic orbits that extend to large distances into the chaotic sea. Both mechanisms are related since the unstable asymptotic curves determine the path followed by the chaotic orbits starting close to them, in order to reach the large chaotic sea.

(2) The unstable asymptotic curve U of the main unstable periodic orbit O (0.5,0) surrounds the two main islands $O_1$ and $O'_1$ forming lobes into the large chaotic sea that become longer with time. From time to time it returns close to the islands $O_1$ and $O'_1$ contributing to the stickiness of the first type around these islands. The same happens for the stable asymptotic curve S of the main orbit O when mapping backwards in time.

Similar asymptotic curves start at the simplest unstable periodic orbit 4/9 around the islands $O_1$ and $O'_1$ that has 9 points ($P_1,…,P_9$) around $O_1$ and another 9 points around $O'_1$. There are two directions of the unstable asymptotic curves ($U_+$ and $U_-$) and of the stable asymptotic curves ($S_+$ and $S_-$) from each point $P_i$. The asymptotic curves $U_+$ and $S_+$ are closer to the last KAM curve of the island $O_1$ and they are called slow because they make more oscillations around the island $O_1$ before escaping to the large chaotic sea, than the outer curves $U_-$, $S_-$ which are called fast. The overall stickiness due to the inner curves is very similar to the stickiness due to the outer curves; their only difference is that the latter case spend more time in the large chaotic sea. The sticky regions due to the asymptotic curves $U_\pm$ are similar to the sticky regions due to the asymptotic curve U from O (see figures 9a, b), because the curves $U_\pm$ cannot intersect the curve U. This explains also the fact that the nonasymptotic orbits of figure 1, that have initial conditions close and outside the last KAM of $O_1$ form sticky regions similar to the ones produced by the asymptotic curves. The same is true for the unstable asymptotic curves of all the periodic orbits that are located close and outside the last KAM of $O_1$.

(3) Stickiness is related to the "stretching numbers" $a$ (i.e. the local "Lyapunov characteristic number" LCN after only one iteration) along the orbits. The average value of $a$ is the "finite time Lyapunov characteristic number" $\chi(t)$. For relatively short times t the finite time LCN is initially smaller for $S_+U_+$ than for $S_-U_-$. This explains why the asymptotic orbits along $S_-U_-$ go faster into the large chaotic sea. However, all the values of $\chi(t)$ tend to the same "Lyapunov characteristic number" at t becomes larger and tends to infinity. The stretching number $a$ varies from successive minima and maxima. The minima correspond to turning points of the curve $U_+$, and produce the maximum stickiness.

(4) We define the "escape time" (or "initial stickiness time") as the time needed by an orbit initially located near and outside the last KAM of $O_1$, to reach the large chaotic sea. We calculated the "escape times" for a uniform grid of initial conditions in a relatively large region outside the island $O_1$. We found that the lines that separate the fast escapers from less fast escapers coincide with various arcs of the stable asymptotic curves of the unstable periodic orbit 4/9. In order to understand the role of the stable asymptotic curves in separating fast and slow escapers we consider orbits starting along two lines close to an arc of $S_+$ reaching the point $P_1$. Orbits starting on the inner side of $S_+$ (along a curve $K^+$, closer to the center of $O_1$) approach $P_1$ and then deviate along and close to the curve $U_+$ ($P_1$) from $P_1$. Later on the curve $K^+$ reaches the neighbourhood of the points $P_9$, $P_8$, $P_7$ and then it goes around the island $O_1$ and escapes downwards to the large chaotic sea. On the other hand the orbits starting along a line $K^-$ outside $S_+$, after reaching the neighbourhood of $P_1$ go close to $U_-$ and after only a few oscillations they escape into the large chaotic sea.

(5) We have calculated quantitatively the overall stickiness around the island $O_1$ by finding the density of the images of points starting at the same grid of points as in the case of escapes. The maxima of density appear close to the asymptotic curves $U_+$, $U_-$ and $U$. Thus the regions of overall stickiness around the island $O_1$ are very different from the escape regions (regions of initial stickiness) in the same area. The overall sticky regions follow the asymptotic curves $U_+$ and $U_-$ while the escape regions follow the asymptotic curves $S_+$ and $S_-$.

(6) As time progresses the maxima of density due to stickiness become less pronounced. Thus we found that beyond the 1000th iteration the density of the points outside the islands tends to be uniform, and no appreciable stickiness appears any more.

(7) The connection between the asymptotic curves of the unstable orbits O and 4/9 is due to the existence of heteroclinic intersections of the manifolds from O and 4/9. However the overall similarity of the aysmptotic curves $U$, $U_+$ and $U_-$ is due to the fact that these curves cannot intersect themselves or each other. Thus they are forced to follow each other, although they differ in the details.

This new type of stickiness studied in the present paper seems to have useful applications. E.g. in a recent paper Voglis et al. (2006) found an application of this new type of stickiness in the formation of spiral arms in barred galaxies.

**Acknowledgements:**
We thank Dr. R. Dvorak for useful discussions and Dr. C. Efthymiopoulos for several discussions and a number of calculations. We thank also the referee for several useful suggestions.

**CAPTIONS FOR FIGURES**

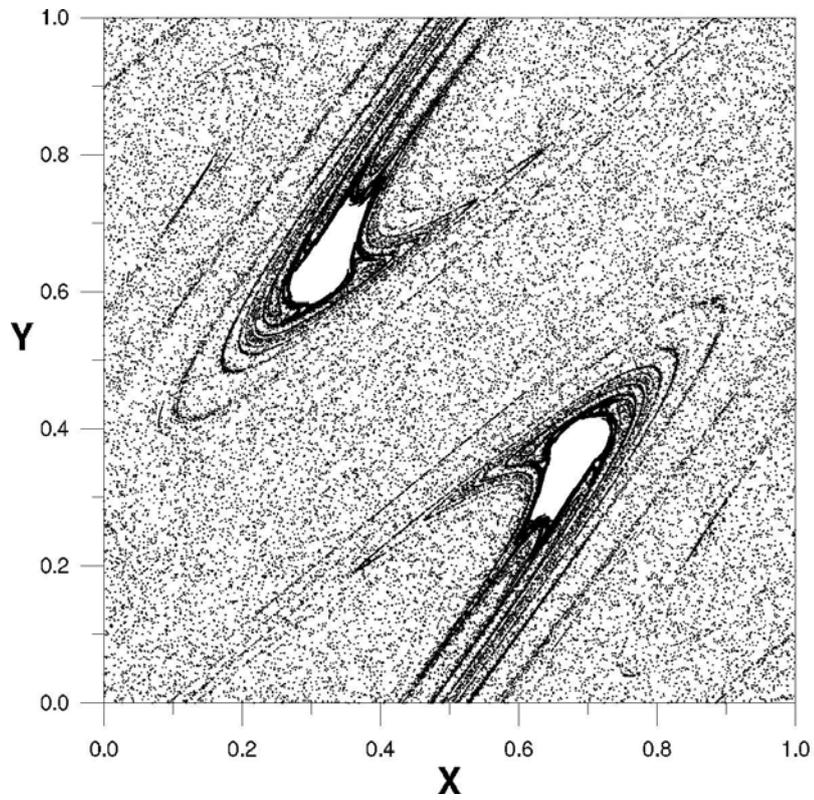

Fig.1     Dark lines in the chaotic sea of the standard map with K=5 and 50 iterations of $10^4$ initial conditions spaced evenly along y=0.3565 with 0.63<x<0.645.

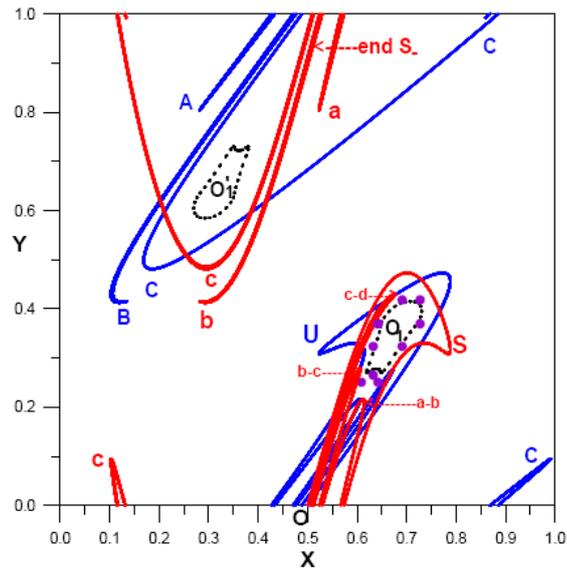

Fig.2    The asymptotic curves U and S from the periodic orbit O (x=0.5, y=0) for K=5. The curves U, S form the lobes A, B, C and a, b, c respectively. They surround the islands around $O_1$ and $O'_1$. The limit of the island $O_1$ is near the orbit 4/9 marked by dots.

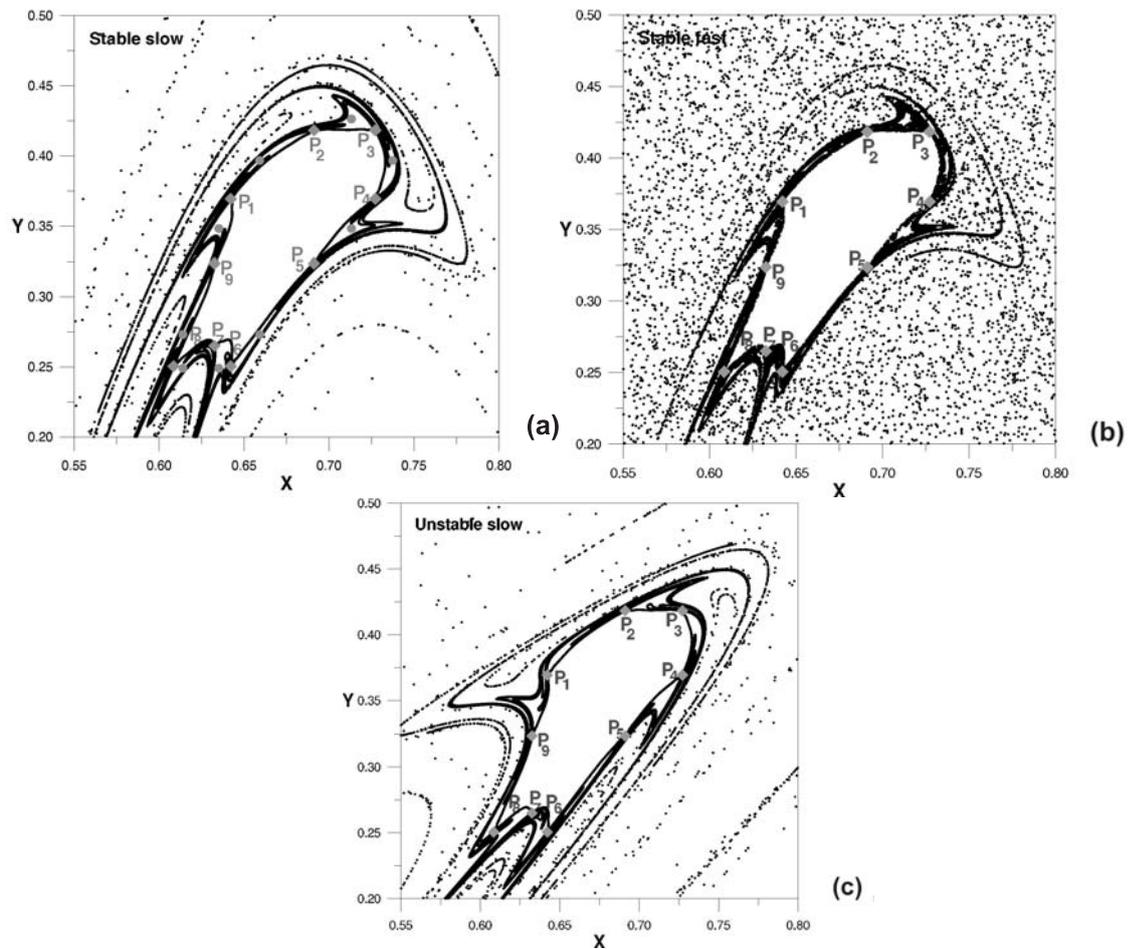

Fig.3      The asymptotic curves from the points $P_1, P_2,…P_9$ of the unstable orbit 4/9 (a) $S_-$ (stable, slow); between the points $P_i$ we mark small islands surrounding the stable periodic orbit 4/9. (b) $S_+$ (stable, fast) (c) $U_-$ (unstable, slow)

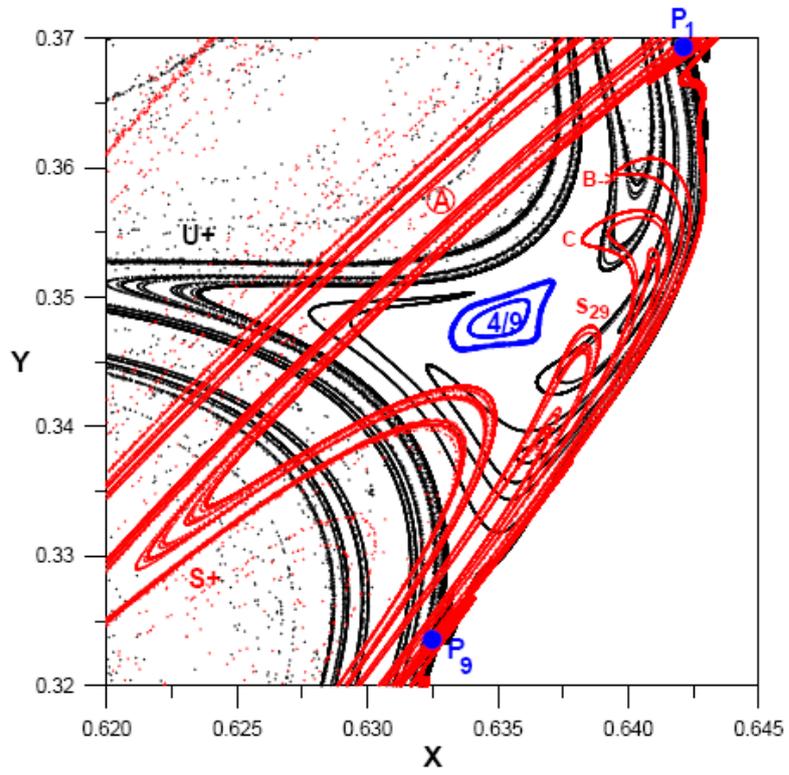

Fig.4　　A comparison of the arcs of $U_+$ and $S_+$. We mark three lobes of $S_+$ (B, C, and S29 from Fig.5b), the open area A of Fig.5b, and the stable island 4/9.

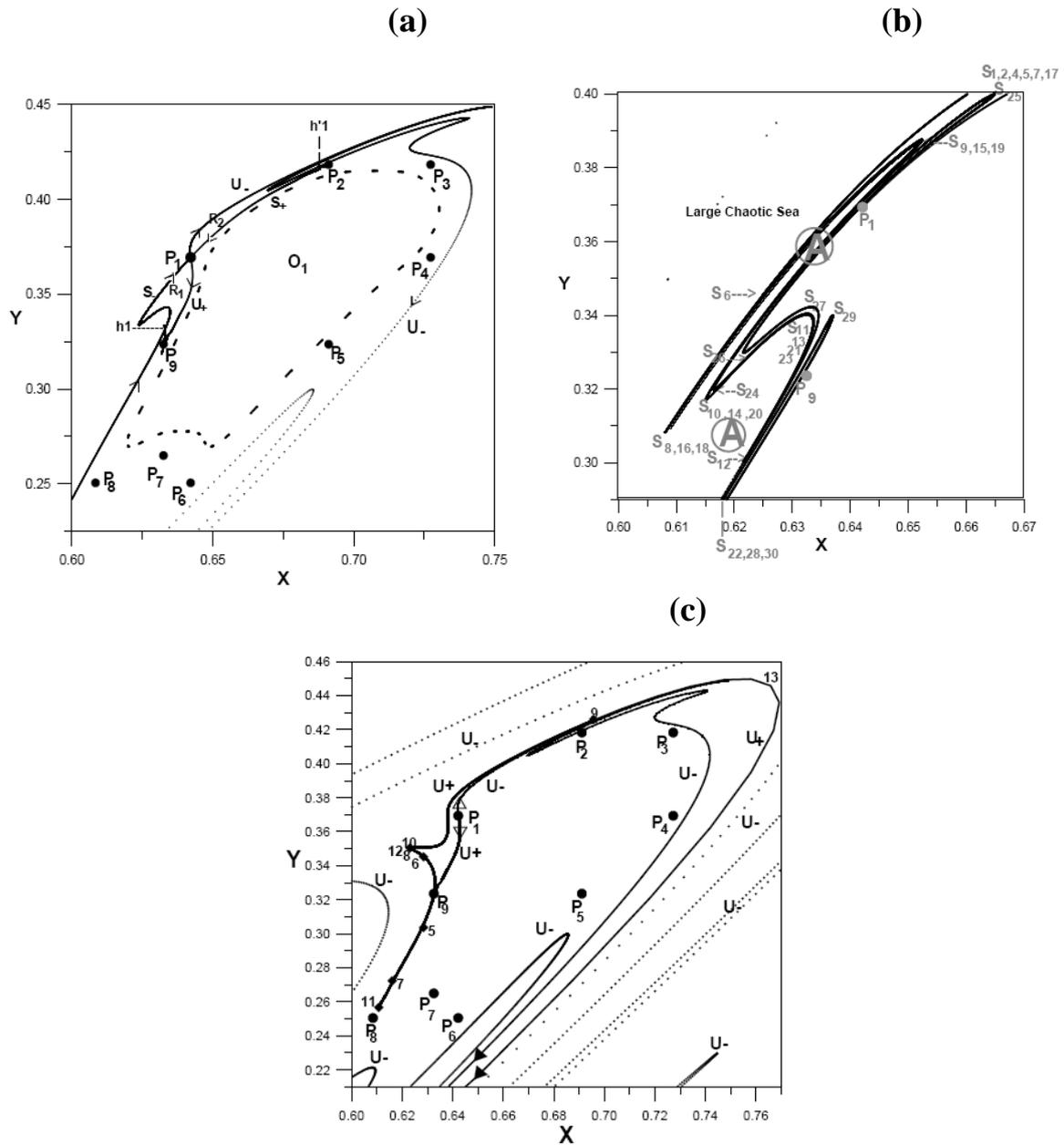

Fig.5 (a) Some details of the asymptotic curves $S_-$, $S_+$, $U_-$, $U_+$ from $P_1$. The dashed curve represents approximately the limits of the island around $O_1$. (b) Successive arcs of the asymptotic curve $S_+$. (c) The asymptotic curves $U_+(P_1)$ and $U_-(P_1)$, from $P_1$. The numbers 5,6,…13 represent maxima or minima y, or x.

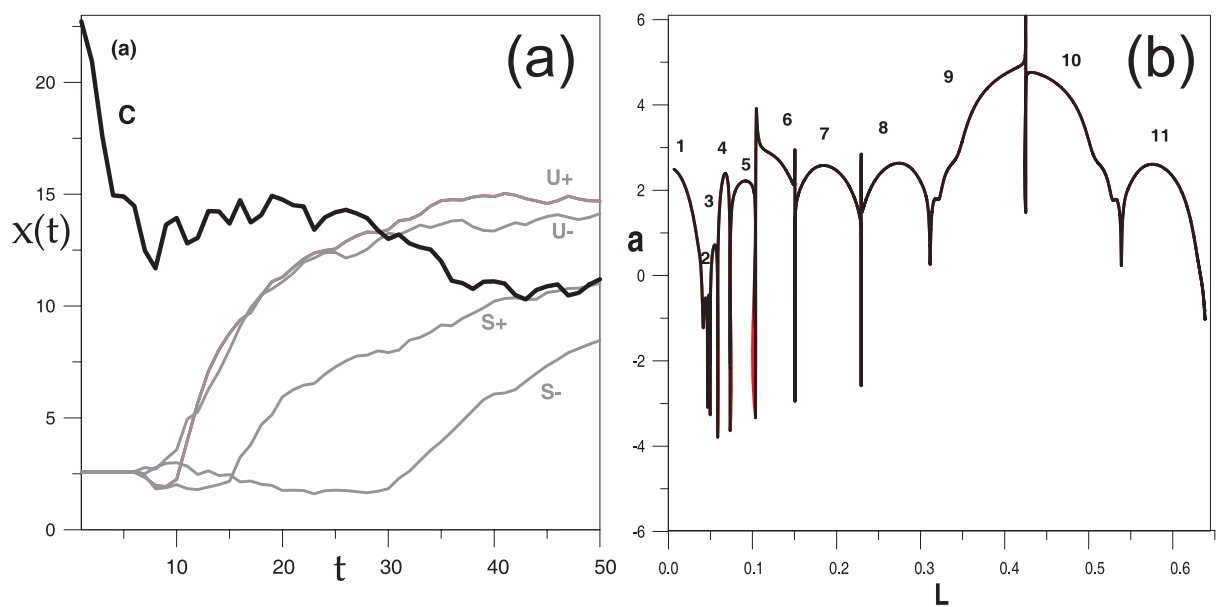

Fig.6 (a) The short time Lyapunov characteristic time $\chi(t)$ as a function of $t$ for orbits along the asymptotic curves $U_+, U_-, S_+, S_-$ (with initial lengths $10^{-9}$, the last two backwards in time) and an orbit C (black line) in the chaotic domain with initial conditions (x=0.3, y=0.5) and initial deviations ($\xi_x = \xi_y = 1$ ). (b) The stretching number $\alpha$ as a function of the distance L from the periodic point $P_1$ along the asymptotic curve $U_+$. The curve $\alpha(L)$ is separated into pieces 1,2,…11, each terminating at a minimum (the first piece terminates at the second minimum).

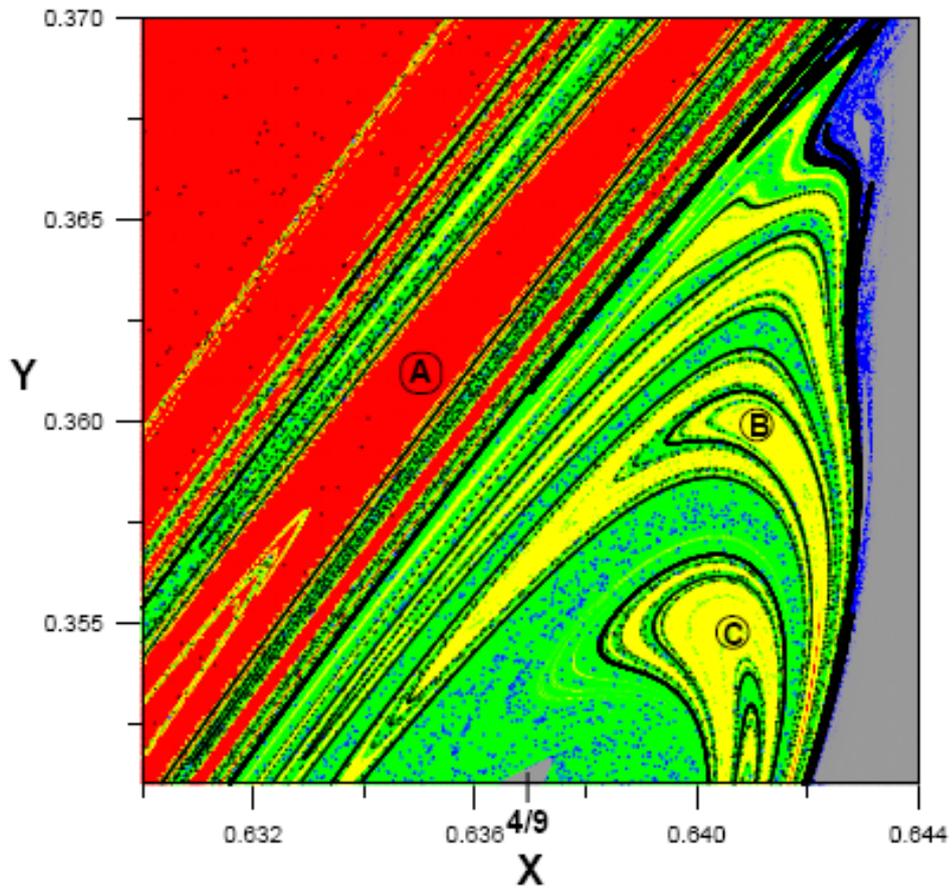

Fig.7  Regions for various escape times: 1-5 (red region A and parallel regions), 5-10 (yellow), 10-100 (green), 100-1000 (blue), >1000 (or non escaping) (gray on the right and close to 4/9). The dark lines represent the stable asymptotic curve $S_+$ of the orbit 4/9.

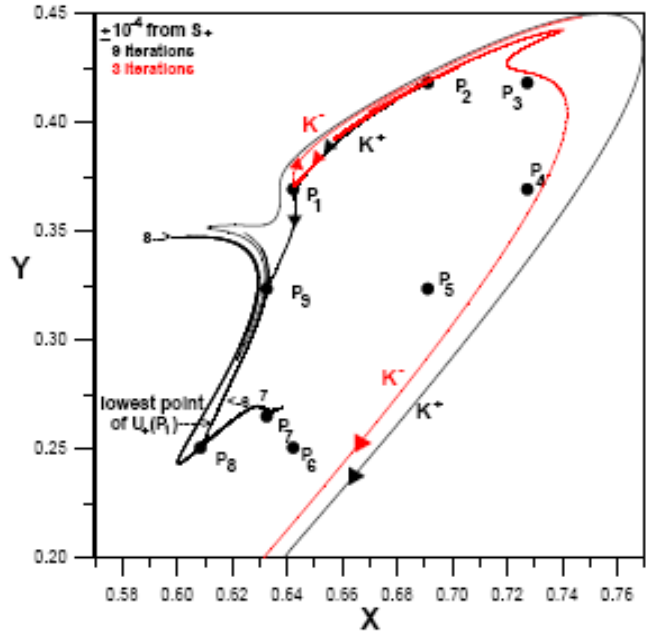

Fig.8  The images of two intervals close and parallel to $S_+$ $K^+$ (black) inside and $K^-$(red) outside $S_+$. The initial interval starts near $P_2$ and ends near $P_1$. The curve $K^+$ continues below $P_1$ close to $U_+$ and after several oscillations it goes around the island $O_1$ into the chaotic sea. The points 6,7,8 mark the ends of the corresponding iterations. The curve $K^-$ continues above $P_1$ close to $U_-$. Its $3^{rd}$ iteration surrounds the island $O_1$ and reaches the chaotic sea.

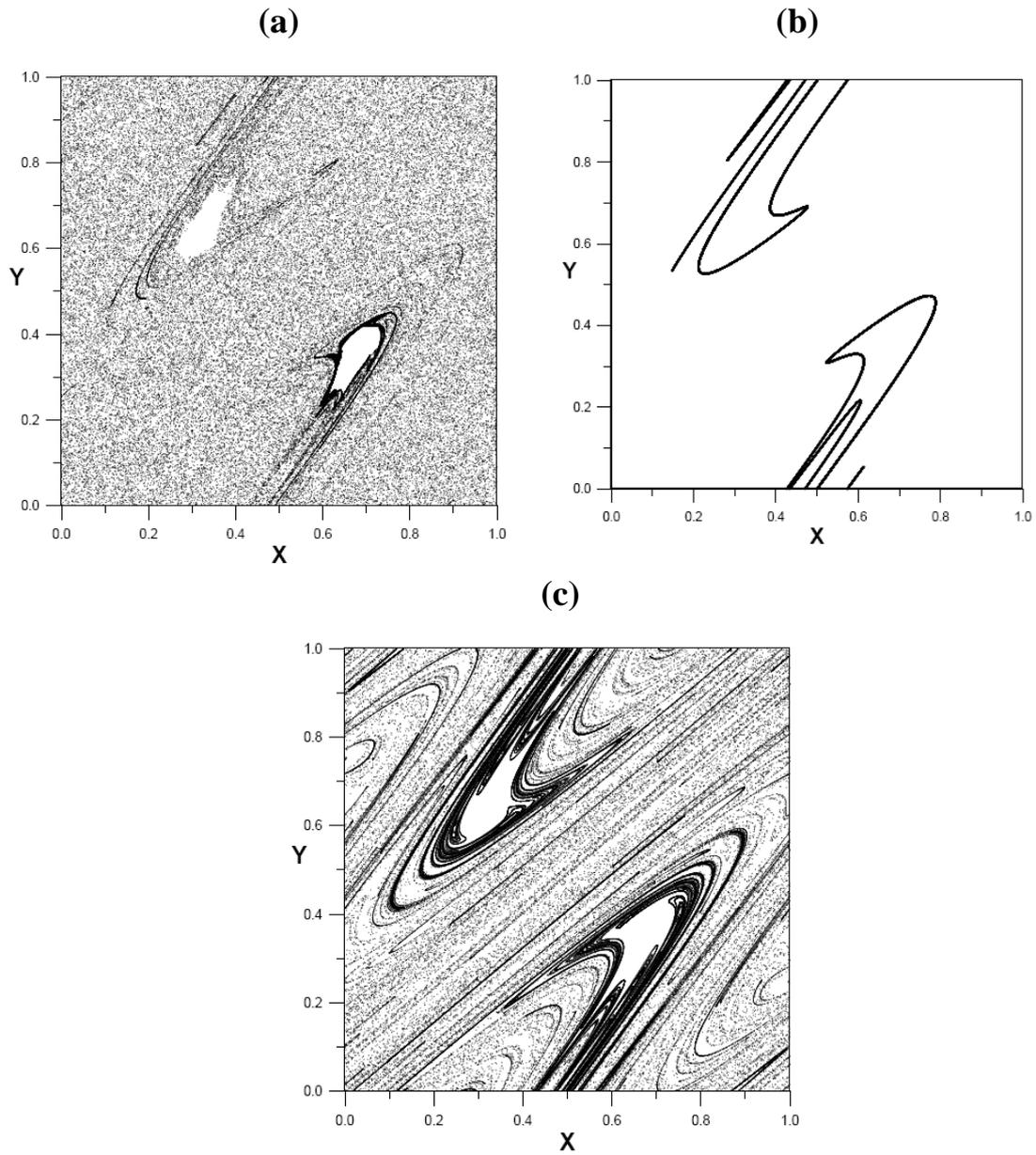

Fig.9　　Sticky regions produced by 500000 points starting along the asymptotic curves (a) $U_+(P_1)$ (15 iterates of 20000 initial points along a $10^{-6}$ initial length from $P_1$) (b) U (15 iterates of 20000 initial points along a $10^{-6}$ initial length from O) (c) U (25 iterates of the same initial points as in Fig. 9b).

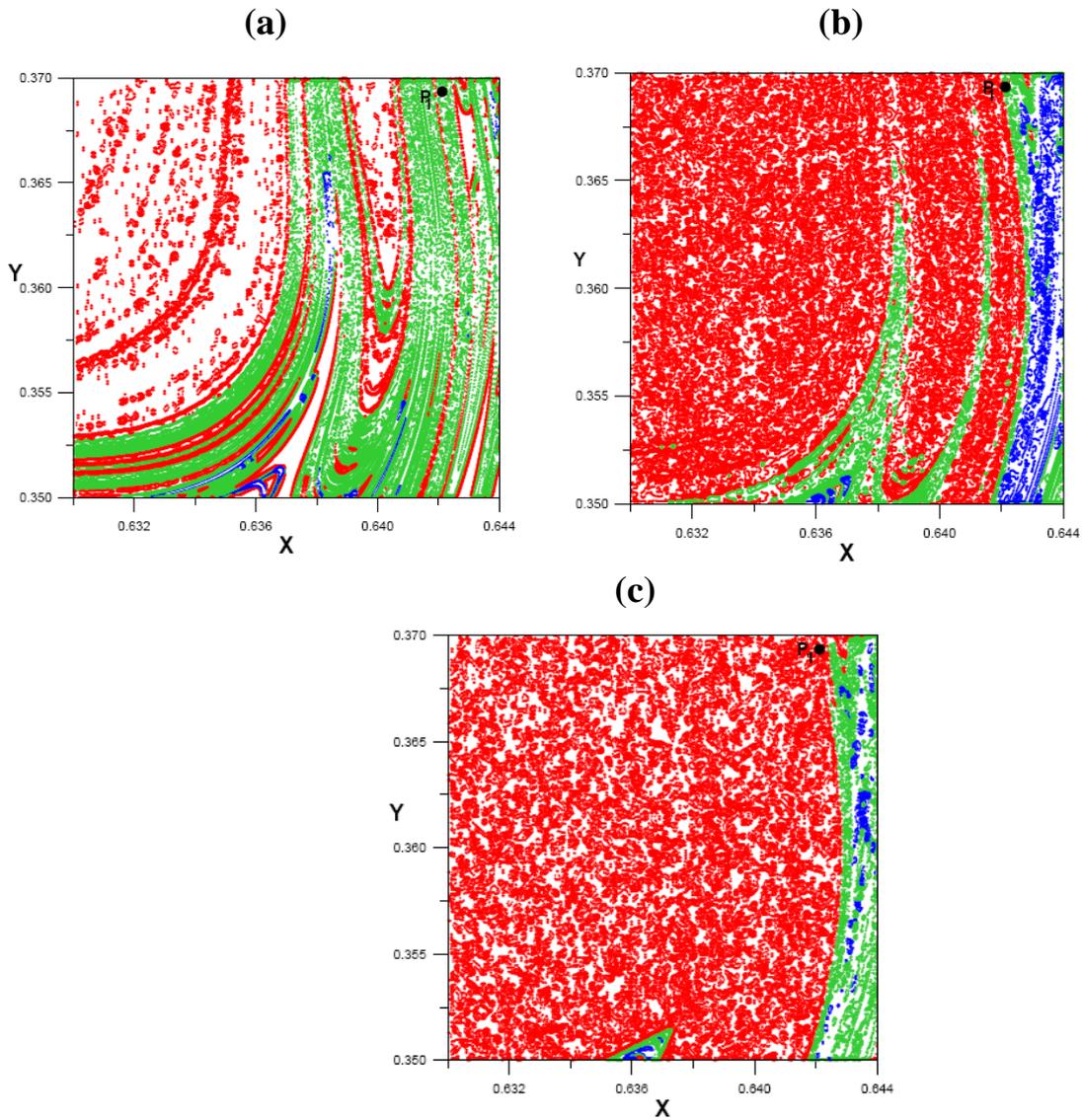

Fig.10   Regions of equal density of 100 iterates from 500000 initial points (50000 equally spaced points with 0.63<x<0.644 along 10 lines of constant y (0.35, 0.352…0.37)). The number of points in a grid 800x800 in the above area is marked with red lines or dots (1-10 points), green (10 points-20points), blue (>20points) and white (empty). (a) iterates 1-100, (b) iterates 400-500, (c) iterates 1000-1100. We mark the position of the point $P_1$.

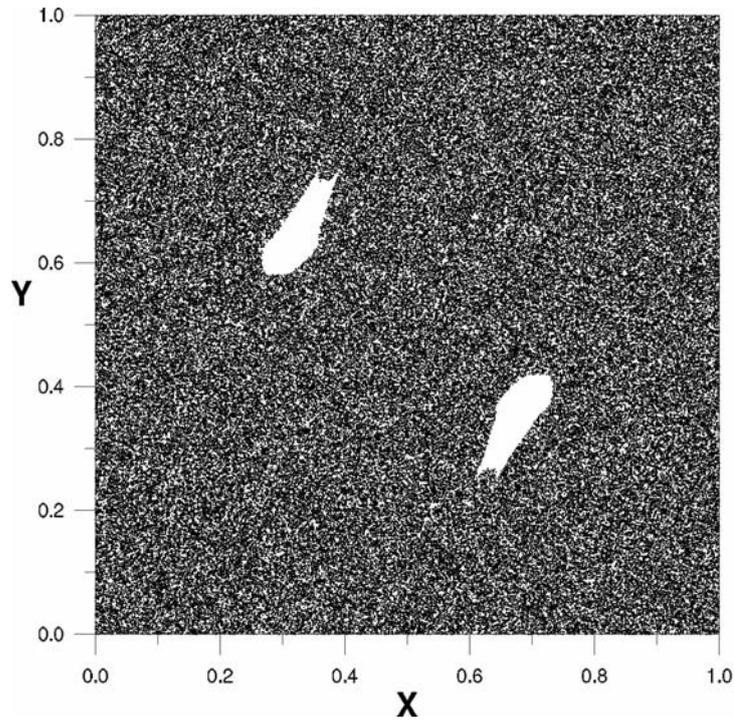

Fig.11  The distribution of 500000 points (50 iterations of the 10000 initial points of Fig.13a) along the asymptotic curve U from O(0.5,0) after a time T=1000, distributed all over the space (0,1)x(0,1).

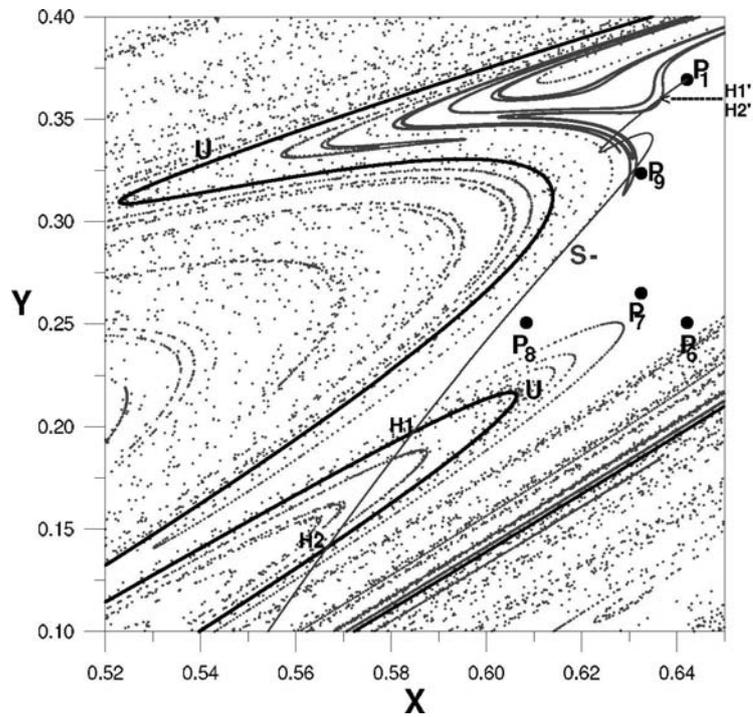

Fig.12  Heteroclinic intersections of the asymptotic curves U and S..